# Enhancing the Electron Mobility in Si-doped (010) β-Ga$_2$O$_3$ films with Low-Temperature Buffer Layers

Arkka Bhattacharyya[1], Carl Peterson[1], Takeki Itoh[1], Saurav Roy[1], Jacqueline Cooke[2], Steve Rebollo[1], Praneeth Ranga[2], Berardi Sensale-Rodriguez[2], and Sriram Krishnamoorthy[1]

[1]Materials Department, University of California Santa Barbara, Santa Barbara, California, 93106, USA

[2]Department of Electrical and Computer Engineering, University of Utah, Salt Lake City, Utah, 84112, USA.

ABSTRACT

We demonstrate a new substrate cleaning and buffer growth scheme in β-Ga$_2$O$_3$ epitaxial thin films using metalorganic vapor phase epitaxy (MOVPE). For the channel structure, a low-temperature (LT, 600 ºC) undoped Ga$_2$O$_3$ buffer was grown followed by a transition layer to a high-temperature (HT, 810 ºC) Si-doped Ga$_2$O$_3$ channel layers without growth interruption. The (010) Ga$_2$O$_3$ Fe-doped substrate cleaning uses solvent cleaning followed by an additional HF (49% in water) treatment for 30 mins before the epilayer growth. This step is shown to compensate the parasitic Si channel at the epilayer-substrate interface that originates from the substrate polishing process or contamination from the ambient. From SIMS analysis, the Si peak atomic density at the substrate interface is found to be several times lower than the Fe atomic density in the substrate – indicating full compensation. The elimination of the parasitic electron channel at the epi-substrate interface was also verified by electrical (capacitance-voltage profiling) measurements. In the LT-grown (600ºC) buffer layers, it is seen that the Fe forward decay tail from the substrate is very sharp with a decay rate of ~ 9 nm/dec. X-Ray off-axis rocking curve ω-scans show very narrow FWHMs, similar to the as-received substrates. These channels show record high electron mobility in the range of 196 - 85 cm$^2$/Vs in unintentionally doped and Si-doped films in the doping range of 2×10$^{16}$ to 1×10$^{20}$ cm$^{-3}$. Si delta-doped channels were also grown utilizing this substrate cleaning and the hybrid LT-buffers. Record high electron Hall mobility of 110 cm$^2$/Vs was measured for sheet charge density of 9.2×10$^{12}$ cm$^{-2}$. This substrate cleaning combined with the LT-buffer scheme shows the potential of designing Si-doped β-Ga$_2$O$_3$ channels with exceptional transport properties for high-performance Ga$_2$O$_3$-based electron devices.

Power electronic systems form the backbone of the modern world's electrical infrastructure. With the world's global energy consumption on the rise, advances in semiconductor power devices are gaining increased importance to develop energy-efficient low-cost power systems that can reduce energy waste. Electric power conversion is required in almost all sectors of the present-day world. The adoption of power devices using wide-bandgap (WBG) semiconductors such as SiC ($E_g$ = 3.2 eV) and GaN ($E_g$ = 3.4 eV) has made high-power solid-state power switch conversion technology very efficient and compact compared to the existing Si-based technologies. Shifting to ultra-wide bandgap (UWBG) semiconductors ($E_g$ > 4 eV) can lead to further enhancements in the system-level size, weight, and power (SWaP) efficiency [1].

$\beta$-$Ga_2O_3$ is an emerging ultra-wide bandgap (UWBG) semiconductor ($E_g$ = 4.6 - 4.9 eV) with tremendous promise to enable power-efficient high voltage power devices and systems. The high projected breakdown field (~8 MV/cm) of $\beta$-$Ga_2O_3$ translates to BFOM (Baliga's Figure of Merit) several times larger than that of WBG semiconductors like SiC and GaN [1], [2]. $\beta$-$Ga_2O_3$ is the only UWBG semiconductor that offers the advantage of producing large-area native bulk substrates from melt-grown techniques – offering potentially lowered costs for large-scale manufacturing at a much higher device yield and uniformity [3]. Although $\beta$-$Ga_2O_3$ device technology is barely a decade old, the achieved device performances to date are already superior and impressive compared to the incumbent power semiconductors [4]–[9]. $\beta$-$Ga_2O_3$-based devices with breakdown voltage over 8 kV and critical breakdown fields exceeding the theoretical limits of SiC and GaN (breakdown field strengths >5 MV/cm) have been demonstrated - establishing $\beta$-$Ga_2O_3$ as the most promising candidate material for next-generation solid-state power-switching applications [8], [10]–[15].

Increasing the channel mobility in $\beta$-$Ga_2O_3$ epitaxial films has been a very active part of $\beta$-$Ga_2O_3$ materials research [16]–[20]. Higher channel mobility is the key to achieving low on-resistances, enabling low conduction and switching losses in devices. Increasing channel mobility in uniformly doped, delta-doped, and modulation-doped $Ga_2O_3$ channels has been demonstrated using MOVPE and MBE growth techniques with great amounts of success [19], [21]–[24]. State-of-the-art mobility values have been reported in MOVPE-grown (010) and (100) $\beta$-$Ga_2O_3$ homoepitaxial thin films [12], [16], [17], [19], [25]. MOVPE-grown $Ga_2O_3$ films with low background unintentional impurities have demonstrated room temperature electron mobility values close to the theoretical maximum of ~ 200 $cm^2$/Vs. MOVPE and MBE-grown $Ga_2O_3$ epilayers have been most widely developed and used for lateral devices. A major issue with developing channels for lateral devices is the presence of a parasitic charge at the epilayer-substrate interface [26], [27]. This residual Si charge is believed to come from the substrate polishing process or ambient contamination [27]. The presence of this charge peak causes bulk-related (buffer) leakages in lateral transistors that can cause premature breakdown and reduce the devices' performance, longevity, and reliability [28], [29]. Additionally, the presence of this undesired charge also makes the study and analysis of electron transport in these layers difficult [30]. Many different techniques to manage this parasitic channel were proposed and demonstrated. Compensation doping at the substrate interface using Fe, Mg delta doping in MBE, Fe and Mg ion implantation, and Mg-doped buffer in MOVPE films were demonstrated [26], [27], [31], [32]. Fe delta doping in MBE leads to a long Fe forward tail and requires the growth of thick buffers > 500 nm to mitigate compensation effects in the channel region [26]. Ion-implantation technique requires high-temperature activation annealing that leads to dopant diffusion issues [27]. Although the techniques to suppress this parasitic channel were very effective, making use of these techniques for multi-kilovolt class device performance have yet to be demonstrated.

In this work we propose a new substrate wet cleaning process for (010) semi-insulating Fe-doped β-$Ga_2O_3$ substrates (Novel Crystal Technology, Japan) that uses a 30 min HF (hydrofluoric acid) treatment. We demonstrate that when this HF cleaning process is used in conjunction with low-temperature (600 °C) grown buffer layers, exceptionally high electron mobilities in uniformly Si-doped β-$Ga_2O_3$ channels can be achieved over a doping range of $10^{16}$ to $10^{20}$ cm$^{-3}$. Enhanced electron mobility in Si delta-doped β-$Ga_2O_3$ channels were also realized. A systematic analysis of the proposed process flow and the results achieved is presented in this work, using atomic force microscopy or AFM (morphological), secondary ion mass spectroscopy or SIMS (elemental), high-resolution X-ray diffraction or HR-XRD (structural), Hall and capacitance-voltage measurements (electrical). Results indicate no evidence of any damage using the HF cleaning step. The device performance of these channels were previously reported. Lateral transistors that utilized these channels demonstrated multi-kilovolt breakdown voltages up to ~ 3kV and power figure of merit of ~ 1 GW/cm$^2$ that show remarkably low reverse leakage [12]. In this work, we have presented a comprehensive material and process flow characterization of the channel stacks.

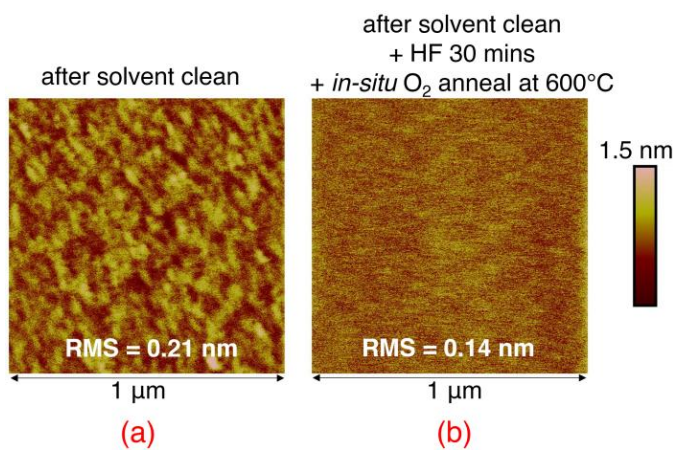

Figure 1: Atomic force microscopy scans (1 × 1 µm$^2$) for the (010) Fe-doped $Ga_2O_3$ substrate before and after the substrate cleaning process. (a) as-received substrate (b) after 30 mins of HF (49%) treatment and an in-situ $O_2$ anneal for 10 mins at 600 °C.

The substrates used for this work are (010) oriented Fe-doped $Ga_2O_3$ substrates acquired from Novel Crystal Technology (NCT), Japan. The as-received substrates were first cleaned sequentially in acetone, methanol, acetone, methanol, and DI water for 5 minutes each in an ultra-sonication bath. The substrates were then submerged in an HF (49%) bath and were kept there for ~ 30 mins. After that, the substrates were transferred to a running DI water bath for a few minutes. The substrates are then blow-dried using an $N_2$ gun and immediately loaded into the MOVPE reactor. The substrate surface was studied before and after the cleaning process using atomic force microscopy (AFM) as shown in Figure 1. The as-received substrates were very smooth to begin with; an RMS roughness of ~ 0.2 nm (Figure 1(a)). Figure 1 (b) shows the surface morphology after the HF treatment and 10 mins of in-situ $O_2$ anneal (details provided later) which was performed inside the MOVPE reactor before epitaxial growth. It can be seen that the substrate surface becomes smoother than the as-received ones. This confirms that the HF cleaning procedure followed by $O_2$ anneal presented here did not result in any morphological degradation of the substrate surface; rather it improves it by making it even more atomically smooth compared to the as-received ones. It is to be noted that the solvent cleaning, HF treatment, and oxygen annealing (all steps) combined result in the substrate cleaning and smoothening effect. The whole process flow mentioned above before epitaxy is critical in suppressing the Si spike at the substrate regrowth interface.

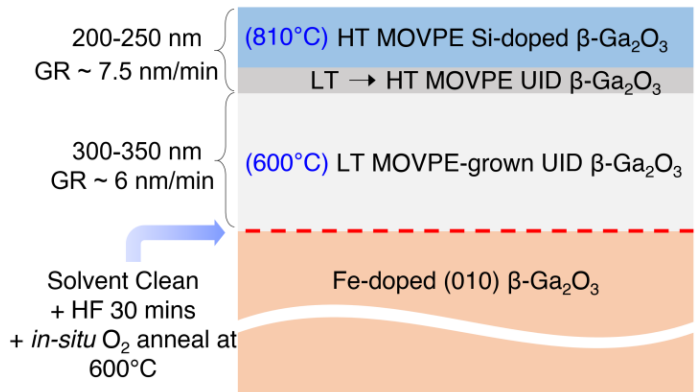

Figure 2: 2D cross-sectional schematic of the channel structure used for the uniformly & intentionally Si-doped films for doping range of $10^{17}$ cm$^{-3}$ to $10^{20}$ cm$^{-3}$. It shows the substrate preparation, LT buffer and the HT Si-doped channel stack used. (GR = growth rate, HT = high-temperature, LT = low-temperature)

The $β-Ga_2O_3$ channels were grown using an Agnitron Agilis' 100 vertical quartz tube reactor which uses TEGa (triethylgallium) and ultra-high purity $O_2$ gases as precursors, diluted silane ($SiH_4$) as the Si source for doping and Ar as the carrier gas. After the substrate preparation, the sample was immediately loaded into the antechamber and transferred to the MOVPE chamber housed in a glovebox. The cross-sectional schematic of the channel-buffer stack is shown in Figure 2. After loading the cleaned substrates, the reactor chamber was immediately pumped down to a chamber pressure of 60 Torr, and the substrates were annealed at 600 °C. For annealing, the substrate temperature was ramped to 600 °C and kept at steady state for 10 mins with Ar:$O_2$ gas flow ratio of 5:4. Following this, the undoped buffer was also grown at 600 °C using only TEGa and $O_2$ gas at a growth rate of ~ 6 nm/min [25]. The undoped buffer thickness is kept around 300 - 350 nm. After this, the substrate temperature was ramped to 810 °C without any growth interruption. While the temperature is ramping, a transition layer of an un-doped $Ga_2O_3$ is grown which is estimated to be around 15-20 nm. After this, the Si-doped channel was grown for which the thickness was kept around 200-250 nm for different Si-doping densities which was varied from $10^{17}$ $cm^{-3}$ to $10^{20}$ $cm^{-3}$ by commensurate scaling of the silane molar flows. The high-temperature uniformly doped layers were grown at a growth rate of ~ 7.5 nm/min.

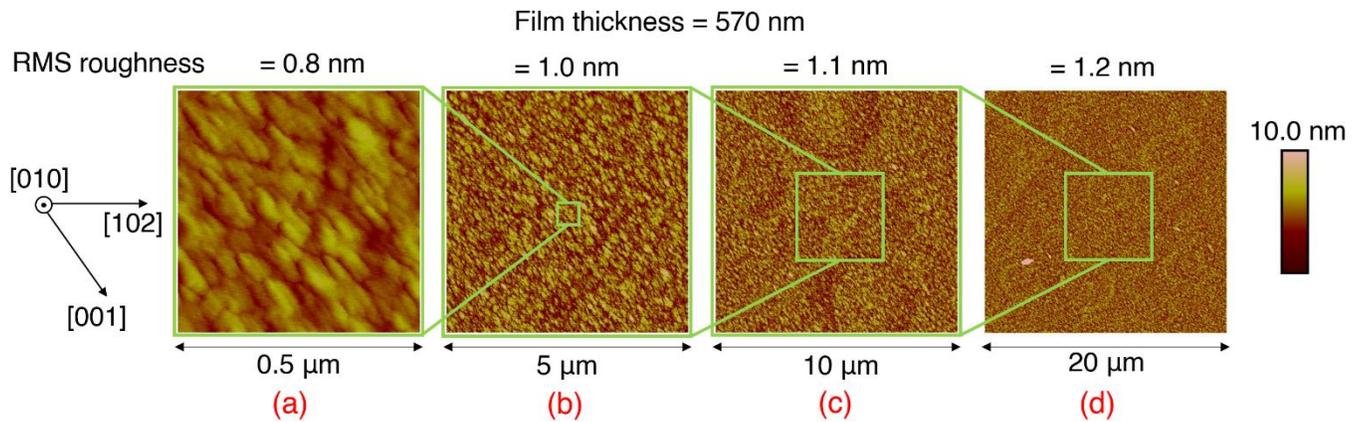

Figure 3: AFM scan of a ~ 570 nm thick epilayer that has 330 nm of LT buffer and 240 nm of Si-doped layer (n = 3.6 ×$10^{17}$ $cm^{-3}$). Small to large area scans showing smooth surface morphology and uniformity. (a) 0.5 × 0.5 µm², (b) 5 × 5 µm², (c) 10 × 10 µm² and (d) 20 × 20 µm².

Figure 3 shows the AFM scans of a sample which has a total target thickness of 570 nm. Figure 3(a) to (d) shows a small area scan of 0.5 × 0.5 µm² to large area scans of 20 × 20 µm². The films show very smooth surface morphologies with atomically flat surfaces with average RMS roughness value of ~ 1nm. These roughness values are comparable or better than previous reports of films grown using MOVPE and MBE for similar film thicknesses and growth rates [16], [33]–[37]. The sub-nanometer RMS roughness indicate that the growth condition used for both 810 °C and 600 °C growths are sufficient to achieve adequate Ga adatom diffusion lengths to maintain an atomically flat surface morphology [25]. Figure 3 (a) small area scan shows the groove-like elongated features that are typically oriented along the [001] in-plane direction. This indicates signs of step-bunching along this direction that has been seen in MBE and MOVPE grown films and is consistently present in high quality Ga$_2$O$_3$ films on (010) oriented substrates [16], [17], [38].

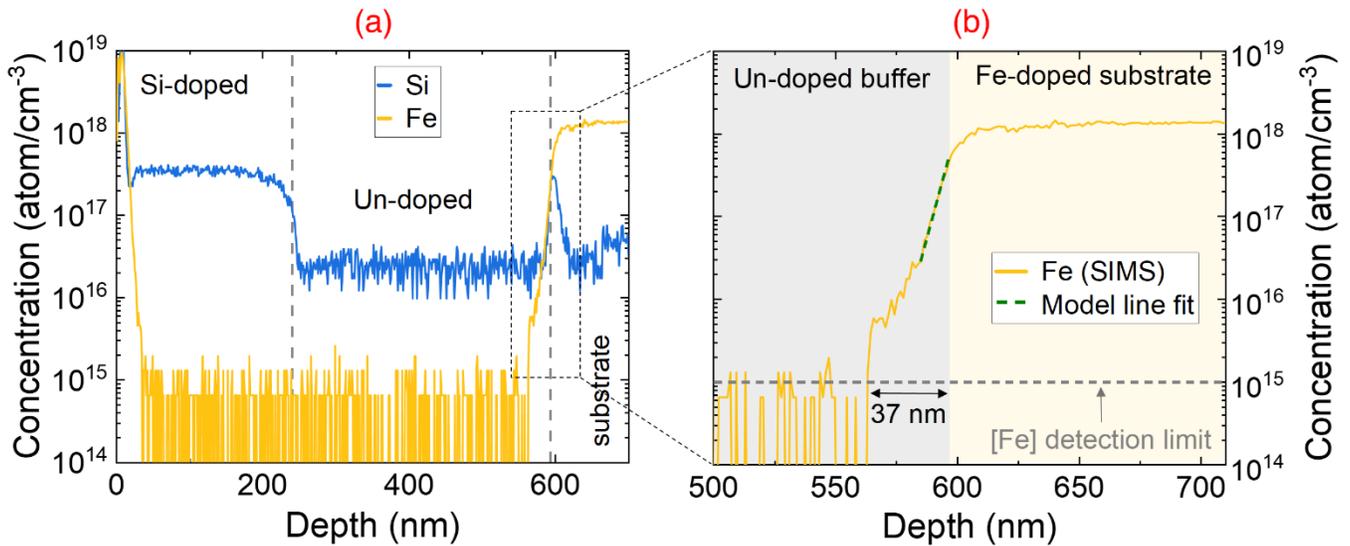

Figure 4: SIMS scan of Si and Fe impurities in a film that is a ~ 570 nm thick epilayer with a 330 nm of LT buffer and 240 nm of Si-doped layer (n = 3.6 ×10$^{17}$ cm$^{-3}$). This is the same sample shown in Figure 3. (a) Atomic concentrations of Fe and Si vs depth of the channel. (b) Magnified scan showing the Fe decay tail from the substrate into the LT buffer.

To quantify the Si and Fe impurity concentrations and dopant profiles, SIMS was performed. SIMS was done using a CAMECA IMS 7F tool. Figure 4 (a) shows the Si and Fe profile from the film surface up to 100 nm into the substrate. First, it can be seen that the Si peak density ([Si]$_{peak}$~ 3×10$^{17}$ cm$^{-3}$) at the epilayer-substrate regrowth interface is less than half of the Fe atomic density ([Fe]$_{substrate}$ ~ 1.5×10$^{18}$ cm$^{-3}$) in the substrate. Also, the Si peak density in Figure 4(a) is 1-2 orders less than the peak Si density reported earlier for studies that used (010) Fe-doped substrates from NCT, Japan, without HF cleaning (typically [Si]$_{peak}$ > 5×10$^{18}$ cm$^{-3}$) [16], [27], [39]. *Feng et al.* used diluted HF dip and got a reduced Si spike density [19]. Based on this observation, any Si parasitic channel is likely compensated by the Fe in the substrate. The LT un-doped buffer showed a background Si concentration of 1-2 ×10$^{16}$ cm$^{-3}$ which is similar to the background values reported earlier in MOVPE thin films [25].

The Fe tail from the substrate into the buffer requires some additional quantitative analysis. The surface segregation of Fe in epitaxial β-Ga$_2$O$_3$ films has been proposed to be the dominant mechanism behind such behavior and studied in earlier reports. The formal analysis uses a model of Fe incorporation and surface segregation at different growth temperatures as reported in reference [40]. The model defines a surface segregation coefficient, R, which assumes that a fraction (R) of dopant atoms (here it is Fe) segregates at the growth surface while a fraction (1-R) incorporates into each monolayer of the film that is growing. Higher the value of the surface segregation coefficient, R, lower will be the dopant incorporation in the preceding monolayers, and longer will be the forward dopant decay tail. Now, from the experimentally measured Fe decay profile, the R value for Fe can be calculated for the growth condition used for the LT-buffer growth. After the growth of 'n' monolayers, the surface coverage after the $n^{th}$ layer and the Fe sheet concentration in that layer is given by,

$$\theta_n = \theta_o R^n$$

$$[Fe]_n = \theta_o(1-R)R^{n-1}$$

where $\theta_o$ is the initial surface coverage, $\theta_n$ and $[Fe]_n$ are the surface coverage and sheet concentration in the '$n^{th}$' monolayer, respectively. Sheet concentration of Fe in the 1st layer is given by

$$[Fe]_1 = \theta_o(1-R)$$

Now, solving for R, one can determined the analytical expression for R as,

$$R = \{[Fe]_n / [Fe]_1\}^{1/(n-1)}$$

Figure 4 (b) shows the line fit of this model to the Fe forward tail. The dopant forward decay rate is defined as the distance required in the growth direction (from the substrate toward epilayer surface) for one order of magnitude change in the dopant concentration. The forward decay rate of Fe from the SIMS data is extracted to be around ~ 9nm/dec. The $\theta_o$ is calculated by integrating the SIMS extracted Fe concentration from the surface to the substrate (the top 50 nm from the film surface in the SIMS data is omitted). The $\theta_o$ at the substrate surface is found to be ~ $1.2 \times 10^{12}$ cm$^{-2}$. Assuming a constant R for the given growth condition and fitting the SIMS profile to an exponential Fe dopant decay, the R value extracted from the Fe surface segregation model in the LT-buffer is 0.948. This value is consistent with sharp Fe decay rate seen from SIMS. Additionally, this value is also consistent with the expected R values discussed in the previous reports of Fe and Si surface segregation observed in MOVPE and MBE grown β-$Ga_2O_3$ epitaxial films [40], [41]. Additionally, it can also be seen that the total film thickness from the substrate interface to the layer where Fe concentration falls below $10^{15}$ cm$^{-3}$ (detection limit) is only about ~ 37 nm, compared to > 300 nm reported in the literature, a 10× improvement [26], [40]. For an initial surface coverage, $\theta_o$ of ~ $1.2 \times 10^{12}$ cm$^{-2}$ and R = 0.948, the film thickness required for the Fe concentration to fall below $10^{15}$ cm$^{-3}$ is calculated to be ~ 42 nm. This value agrees well with the experimentally extracted value. This shows that the Fe impurities are sufficiently far away from the doped channel. This analysis also shows that LT-buffers can be an efficient way to isolate doped channels from Fe impurities and also tailor the Fe dopant decay rate in $Ga_2O_3$ films to very small values. This is very critical in achieving very low compensation in $Ga_2O_3$ epitaxial films as will be discussed later.

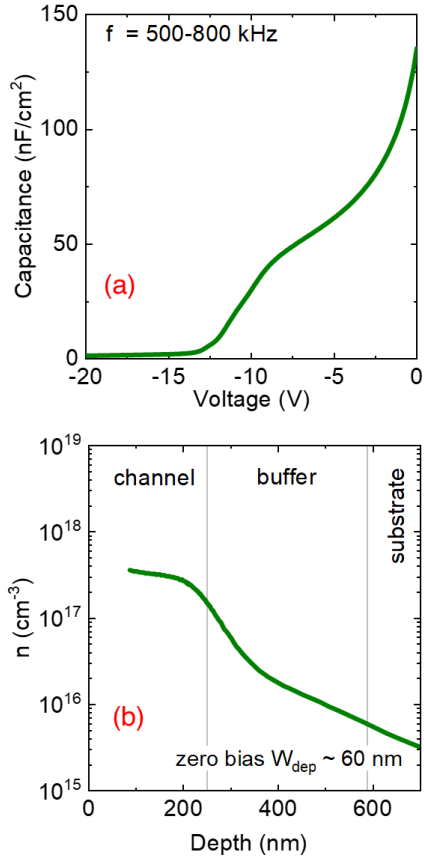

Figure 5: (a) Capacitance-Voltage profile measured on a lateral Ni Schottky diode from which the charge profile was extracted. The sample is a ~ 570 nm thick epilayer with a 330 nm of LT buffer and 240 nm of Si-doped layer (n = 3.6 ×10$^{17}$ cm$^{-3}$). This is the same sample shown in Figure 3 & 4. (b) Apparent charge profile of the channel with LT buffer showing no charge peak at the substrate regrowth interface. The extracted charge density is in agreement with the SIMS-measured Si density in Figure 4(a).

Figure 5 shows the CV profile and the apparent charge profile of this stack. It can be seen the electron density extracted from CV in the doped region matches well with the Si atomic density from SIMS which is ~ 3.6 × 10$^{17}$ cm$^{-3}$ for both (Figure 4(a) and Figure 5(b)). These values also matched well with Hall measurements which will be discussed later. This indicates full Si (~ 100%) activation in the HT Si-doped channels. From the CV profile and the extracted apparent charge profile, it can be seen that the buffer is depleted and there is no charge peak at the substrate regrowth interface. This corroborates well with the SIMS data which explains how the Si is compensated by the Fe in the substrate. Without the Si peak, the Fermi level ($E_F$) at the buffer-substrate interface will get pinned at the Fe trap level (~ 0.8 eV below the conduction band minima) [42]. Hence, with a background doping of ~1 ×10$^{16}$ cm$^{-3}$ in the LT UID buffer, the substrate backside barrier of 0.8eV can deplete a ~ 300 nm of the buffer. This explains how the substrate back-depletion depletes the buffer. Hence, without the Si parasitic channel, all of the conductivity in these films can be attributed to the HT Si-doped region only, as seen from the single bump in the CV profile before pinch-off.

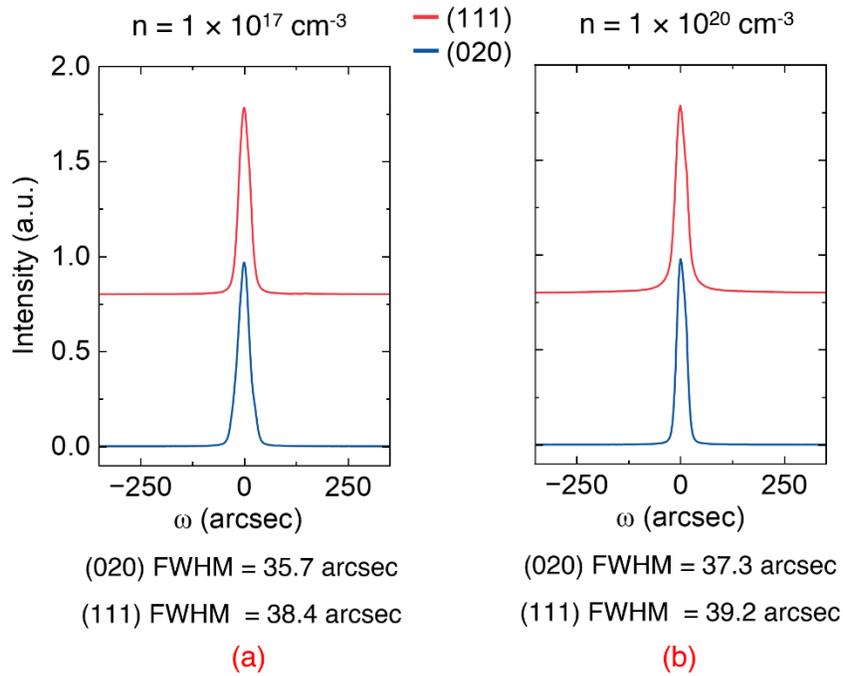

Figure 6: On-axis and off-axis X-ray rocking curve ω-scans on films with Si-doped channels and LT buffer. Shows X-Ray rocking curves for (a) film with a lightly-doped (n ~ 1×10$^{17}$ cm$^{-3}$) channel and (b) a film with heavily doped (n ~ 1×10$^{20}$ cm$^{-3}$) channel region.

Figure 6 shows the X-Ray rocking curve (RC) measurements on these films with LT-buffer layers. On-axis (020) and off-axis (111) plane peaks on different samples with different doping densities were measured. As shown in Figure 6, both low-doped (10$^{17}$ cm$^{-3}$) and heavily doped (10$^{20}$ cm$^{-3}$) showed similar off-axis (111) narrow RC peaks which match closely with the on-axis (020) peaks [18]. The full-width half maximum (FWHM) of the off-axis RC curves were around 35 – 40 arcsec for the samples along the full doping (4 orders) range which nominally were identical to the on-axis FWHM values. This indicates excellent structural quality of these MOVPE-grown films.

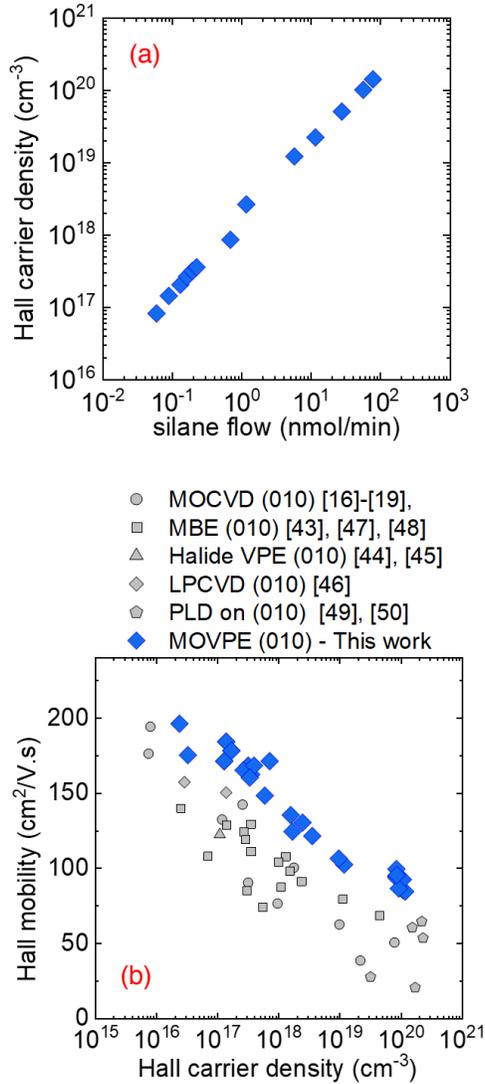

Figure 7: (a) Measured electron density as a function of silane molar flow measured using Hall measurements. (b) RT Hall mobility of the channel stacks as a function of Hall carrier density benchmarked with various state-of-the-art reports [16]-[19], [43]-[50].

Si-doping was achieved in these films by flowing diluted silane (SiH$_4$) into the chamber during the growth of the doped channel region grown at 810°C and doping in the range of ~ $2\times10^{16}$ cm$^{-3}$ to $1.2\times10^{20}$ cm$^{-3}$ was achieved by scaling the silane molar flow rate. The electron concentrations were measured by room-temperature Hall measurements (Ecopia HMS 3000 and HMS 7000). Ohmic contacts were formed by depositing Ti (20 nm)/Au (200 nm) on the four corners of the square samples by using low-power DC sputtering. The electron concentrations and doping profiles were also verified using C-V measurements whenever feasible and applicable. Figure 7(a) shows the measured electron concentrations in these films as a function of silane molar flow rate. It shows that the electron concentrations correlate well with the silane flow scaling and scales linearly with silane flow values. For the low doping densities between $2\times10^{16}$ cm$^{-3}$ and $6\times10^{17}$ cm$^{-3}$, CV was performed using Ni/Au Schottky CV pads which showed excellent agreement with the Hall data. From SIMS, CV, and Hall measurements, it was observed that the Si activation efficiency for the reported doping range of $10^{16}$ to $10^{20}$ cm$^{-3}$ is ~ 100% which also corroborates well with our previous report [25]. This shows that this stack design and growth conditions can be efficiently used to achieve over 4 order swing in the doping densities. Figure 7 (b) shows the measured Hall electron mobility values in these channels that were measured over a large number of samples. As mentioned earlier from CV measurements, the total conductivity in these stacks can be attributed to only the doped channel region. This is further corroborated by depth-dependent field-effect mobility characterization reported earlier [12]. The highest mobility of 196 cm$^2$/Vs (n = $2.3\times10^{16}$ cm$^{-3}$) was measured in a 1.5 µm thick UID film grown at 810°C with 300 nm of LT UID buffer. The UID film had to be grown thicker to have a sufficient sheet charge of ≥ $3\times10^{12}$ cm$^{-2}$ so that reliable Hall measurements can be performed. Mobility of 184 cm$^2$/Vs was measured in a 250 nm thick $1\times10^{17}$ cm$^{-3}$ Si-doped film. Mobility values > 100 cm$^2$/Vs are measured in films with electron concentrations between $1\times10^{18}$ cm$^{-3}$ and $1\times10^{19}$ cm$^{-3}$. For an electron density of $1.2\times10^{20}$ cm$^{-3}$, mobility values as high as 85 cm$^2$/vs were measured. These electron mobility values are the highest to ever be reported in β-Ga$_2$O$_3$ epitaxial films [16]–[19], [43]–[50].

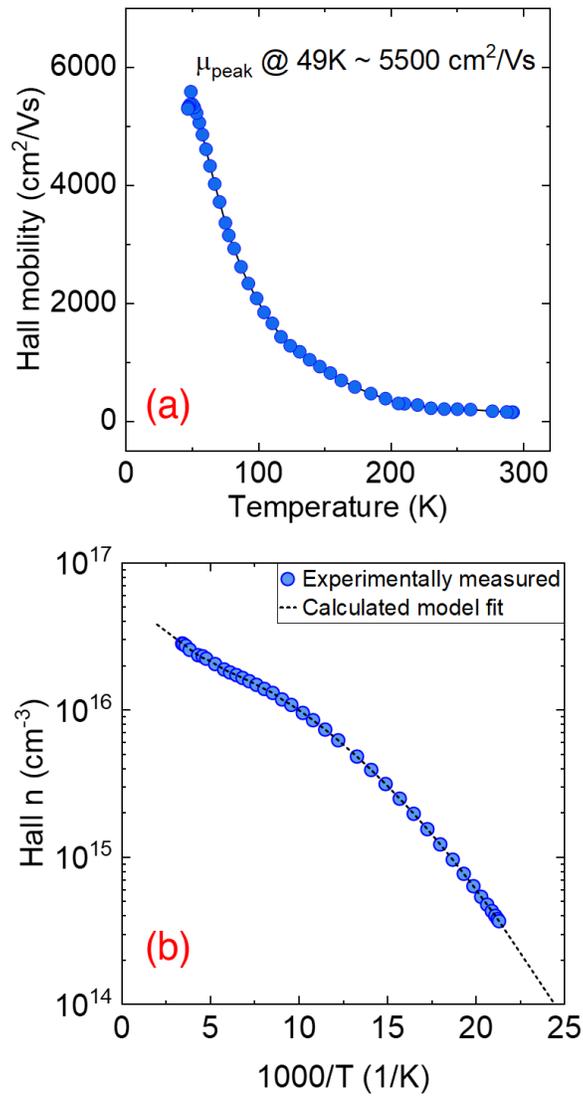

Figure 8: Temperature-dependent Hall measurements on a 1.5 μm thick UID film grown at 810°C with 300 nm of LT UID buffer. (a) Experimental Hall mobility measured as a function of temperature (b) Hall carrier density as a function of 1000/T. Data points represent experimentally measured values. Dashed line represent the calculated charge density using the three donor and a single acceptor charge neutrality model.

To further quantify the compensation and the Si donor behavior in the doped region of the channel, temperature-dependent Hall measurements (40K – 300K) were performed using a Lakeshore Cryotronics 7504 Hall effect system. At room temperature, the electron mobility in a lightly doped β-$Ga_2O_3$ film is dominantly limited by polar-optical phonon scattering and not by ionized impurity scattering. At lower temperatures, scattering from ionized impurities in these films dominate the electron scattering mechanisms that can be leveraged to quantify them. The low-temperature Hall measurement was performed on an un-doped or unintentionally doped film to probe the peak electron mobility and the intrinsic background compensation. For a 1.5 μm thick UID film grown at 810°C with 300 nm of LT UID buffer, the peak electron mobility measured is ~ 5500 cm$^2$/Vs at a temperature of 49K. UID films with better low-temperature peak mobility up to 23,000 cm$^2$/Vs (film thickness of 5 µm) have been reported in MOVPE films grown using trimethygallium (TMGa) precursor, and 11,400 cm$^2$/Vs (film thickness of 3.2 µm) using triethygallium (TEGa) [17], [30]. It is to be noted these films in the literature reports have lower background Si impurity. The peak LT mobility in our UID films is expected to improve by lowering the background doping as it is expected to be limited by ionized impurity scattering. The measured temperature-dependent Hall electron densities were fit using a charge neutrality equation [18]. The best fit to the data was achieved using three donor states and a single compensating acceptor level as shown below,

$$n + N_a = \frac{N_{d1}}{1 + 2e^{-\frac{E_{d1}-E_f}{k_BT}}} + \frac{N_{d2}}{1 + 2e^{-\frac{E_{d2}-E_f}{k_BT}}} + \frac{N_{d3}}{1 + 2e^{-\frac{E_{d3}-E_f}{k_BT}}}$$

where, $n$ and $N_a$ represent the free electron density and the concentration of compensating acceptors, respectively. Concentrations of the three donors are represented by $N_{d1}$, $N_{d2}$, and $N_{d3}$ corresponding to donor energies of $E_{d1}$, $E_{d2}$, and $E_{d3}$, respectively. $E_f$ is the Fermi level at the measurement temperature (T). $E_f$ is estimated by $n = N_c \times e^{-(E_c-E_f)/k_BT}$, where $E_c$, $N_c$, $k_B$ are the energy of the conduction band edge, effective density of state in the conduction band at temperature T, and the Boltzmann constant, respectively. $N_c$ is estimated analytically using an electron effective mass of m*=0.313×m$_o$, and $N_{d1}$, $N_{d2}$, $N_{d3}$, $E_{d1}$, $E_{d2}$, $E_{d3}$, and $N_a$ are the free parameters extracted by curve fitting. The values extracted from the charge neutrality fit is shown in Table I. The activation energies of the three donor levels identified from the best fit are 31.5 meV, 56 meV and 120 meV. Majority of the low and room temperature (over 80%) of the electron density is contributed from the shallow level of 31.5 meV. This shallow donor level is the typical Si donor level expected from theory which is also experimentally verified by several reports [16], [17], [42], [51]. The other two deeper donor levels have also been previously reported but their origin is yet to be identified. Possible sources could be Si occupying octahedral Ga sites, interstitials, antisites, etc. [16], [18]. It was also hypothesized by *Alema et al.* that these deeper level are probed because of the presence of the parasitic Si channel at the substrate interface [30]. But in our case, this parasitic channel is fully compensated and electrically inactive. So, these levels are not originating from this parasitic channel and could be present within the doped epitaxial layer. Therefore, the origin of these deeper levels will require a more detailed investigation in future.

The compensating acceptor concentration extracted is around $N_a$ = 5×10$^{14}$ cm$^{-3}$ which is two orders lower than the electron density. This corresponds to < 2% compensation in these films. It is to be noted that obtaining a very small Fe tail into the epitaxial layers was also one of the key contributing factors that resulted in such low compensation in these films. This is very encouraging because it indicates that by lowering the background electron density even further, very pristine low-doped films ($10^{15}$ cm$^{-3}$) can be achieved with very low compensation and excellent transport properties – ideal for ultra-high kilovolt devices. However, the elucidation of the origin of the background impurities in MOVPE β-Ga$_2$O$_3$ films still remains a challenge.

**Table I. Free parameter values extracted by fitting the three donor and single acceptor charge model to the experimental values. $N_{d1}$, $N_{d2}$ and $N_{d3}$ are the concentrations of the three donors corresponding to the donor energies of $E_{d1}$, $E_{d2}$ and $E_{d3}$, respectively. $N_a$ is the concentration of compensating acceptors.**

| $N_{d1}$ (cm$^{-3}$) | $N_{d2}$ (cm$^{-3}$) | $N_{d3}$ (cm$^{-3}$) | $E_c-E_{d1}$ (meV) | $E_c-E_{d2}$ (meV) | $E_c-E_{d3}$ (meV) | $N_a$ (cm$^{-3}$) |
|---|---|---|---|---|---|---|
| $2.6 \times 10^{16}$ | $3.8 \times 10^{15}$ | $1.4 \times 10^{16}$ | 31.5 | 56 | 120 | $5 \times 10^{14}$ |

The device performance of these channels were reported earlier. We demonstrated lateral transistors utilizing these uniformly Si-doped channels with LT buffers which exhibit state-of-the-art device performance [12]. The transistors showed very low reverse leakage for breakdown voltages up to ~ 3kV. Also, from the transistor transfer characteristics, the threshold voltage matched well with the C-V pinch-off voltages. This clearly confirms the elimination of the parasitic electron channel at the substrate interface using the technique proposed in this work. Moreover, due to enhanced electron mobility, these devices were able to exhibit low on-state resistances as well. In conjunction with effective electric-field management, these devices were able to deliver a high power figure of merit of ~ 1 GW/cm$^2$, setting a new record for Ga$_2$O$_3$ transistor device technology.

To summarize the discussions above, the reason behind such high mobility in these channels can be attributed to the following factors combined: 1) Elimination of the low-mobility parasitic electron channel at the substrate interface (HF cleaning), 2) Minimizing Fe surface riding to the doped channel by using LT buffer (600°C buffer growth), 3) Utilizing substrate back-depletion to electrically isolate the doped channel (300 nm thick buffer), 4) Very low compensation of < 2% ( $N_a$ ~ 5 ×10$^{14}$ cm$^{-3}$), 5) Very high structural quality of the epilayers (off-axis XRC FWHM of 35 - 40 arcsec).

The next part focuses on the Si-delta doping in MOVPE-grown Ga$_2$O$_3$ films. We have previously reported Si-delta doping in Ga$_2$O$_3$ using MOVPE [24], [41], [52]. In this work, we integrate the substrate cleaning and the LT buffer design with Si delta-doping and characterize the electron transport in these films using Hall measurements. Delta doping of β-Ga$_2$O$_3$ using MOVPE was achieved through a growth interruption-based process where the flow of TEGa was stopped and silane was flowed into the chamber. A pre- and post-purge step (30 s each) was performed to remove any unreacted TEGa precursors from the growth chamber before silane was flowed into the chamber. For achieving delta doping, silane was supplied to the chamber under an oxygen ambient. The silicon delta sheet density was controlled by tuning silane flow, silane time, and pre- and post-purge steps. Further details of the process can be found here [41], [52]. The substrates used were the same ones used earlier that underwent the same substrate cleaning, in-situ annealing steps before the growth. The cross-sectional schematic of the whole stack is shown in Figure 9. It contains two regions – the LT buffer and the delta-doped region. The LT buffer is similar to the one used earlier. A 300 nm UID buffer was grown at a temperature of 600 ºC at a growth rate of ~ 6 nm/min. The delta-doped region contains unintentionally doped (UID) $\beta$-$Ga_2O_3$ layers before and after (cap) the delta-doped layer. For the delta-doped region, the UID sandwich layers were grown at a slower growth rate of ~ 3 nm/min for improved thickness control and uniformity. The UID cap thickness after the Si delta-doping was kept at 25 nm. There was no growth interruption in the whole stack other than the Si delta-doping step. We have shown previously that Si delta-doping charge density and profile are a strong function of growth temperature. So the delta-stack and doping were done at two temperatures of 600 ºC and 810 ºC [41]. The growth rates for the delta-doped region for both the growth temperatures were kept the same.

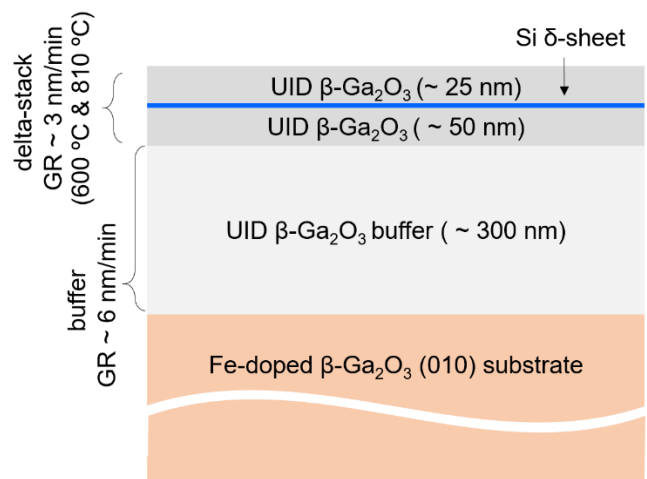

Figure 9: 2D cross-sectional schematic of the delta-doped channel. It shows the two-different growth rates for the LT-buffer and the delta-doped regions. (GR = growth rate)

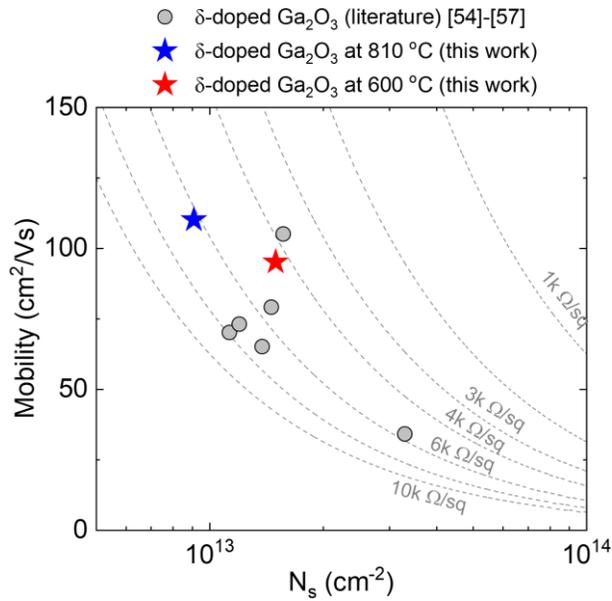

Figure 10: Dependence of the electron mobilities on the sheet charge density for delta-doped $Ga_2O_3$ thin films [54]-[57]. The constant sheet resistance contour plots are shown by dashed lines.

The target delta sheet density was around $1\times10^{13}$ cm$^{-2}$. The sheet charge density ($n_s$) and the mobility of the channel regions were measured at room temperature using Hall measurements. The delta stack grown at 600°C had a sheet charge density of ~ $1.5\times10^{13}$ cm and mobility of 94 cm$^2$/Vs. The delta-stack grown at 810°C gave an $n_s$ value of $9.1\times10^{12}$ cm$^{-2}$ and mobility of 110 cm$^2$/Vs. The higher charge density in the 600°C grown delta-stack is because of higher Si incorporation efficiency at lower growth temperatures for the same growth rates, chamber pressure and silane molar flow rates [41], [53]. Figure 10 benchmarks the mobility as a function of sheet charge density reported in delta-doped $Ga_2O_3$ channels. It shows that the mobility value achieved in this work is a record high for any delta-doped $Ga_2O_3$ channels reported to date [22], [54]–[57]. Hence, the LT buffer scheme can be used to grow uniformly doped channels as well as 2D channels with enhanced electron mobility.

In conclusion, a new substrate cleaning and a hybrid channel structure with low-temperature buffer layers are demonstrated using MOVPE that exhibits record high electron mobility in uniformly Si-doped (3D) as well as delta-doped (2D) $Ga_2O_3$ epitaxial films grown on (010) Fe-doped bulk substrates. For the channel structure, a low-temperature (LT = 600°C) undoped $Ga_2O_3$ buffer is grown followed by transition layers to a high-temperature (HT = 810°C) Si-doped $Ga_2O_3$ channel layers without growth interruption. The substrate uses solvent cleaning followed by an additional HF (49% in water) treatment for 30 mins and in-situ $O_2$ anneal before the epilayer growth that compensates the parasitic Si channel at the epilayer-substrate interface. SIMS analysis shows full compensation of Si peak density at the substrate regrowth interface by the Fe density in the substrate. CV profiling shows no charge peak at the substrate interface, thus, corroborating the observation from SIMS. LT-buffers are very effective in blocking the Fe surface riding into the doped channel, with Fe forward decay rate as sharp as 9 nm/dec. Room-temperature Hall mobilities of 196 cm$^2$/Vs to 85 cm$^2$/vs are achieved in films with electron densities in the range of 2×10$^{16}$ cm$^{-3}$ to 1×10$^{20}$ cm$^{-3}$, respectively. Delta-doped channels with sheet charge of 9.1×10$^{12}$ cm$^{-2}$ and mobility of 110 cm$^2$/Vs were also realized. Transistors based on this structure have shown record high power figures of merit and breakdown voltages up to ~ 3 kV [12]. The transistor characteristics show that the proposed design has excellent leakage performance for multi-kilovolt class devices. The proposed design and process flow show the potential of designing Si-doped Ga$_2$O$_3$ channels with achievable electron mobilities very close to the predicted theoretical maximum which becomes essential as Ga$_2$O$_3$ device technology gears up for the next generation high power and RF applications.

## ACKNOWLEDGMENTS

This work was supported in part by the Air Force Office of Scientific Research under Award FA9550-21-0078 and FA9550-18-1-0479 (Program Manager: Dr. Ali Sayir) and Lawrence Livermore National Laboratory and in part by the II-VI Foundation Block Gift Program. This work was performed in part at the Utah Nanofab sponsored by the College of Engineering, Office of the Vice President for Research, and the Utah Science Technology and Research (USTAR) initiative of the State of Utah. The author(s) appreciate the support of the staff and facilities that made this work possible. The authors also thank Prof. Mike Scarpulla at ECE Department, University of Utah and Prof. Luisa Whittaker Brooks at Chemistry Department, University of Utah for access to electrical characterization tools (parameter analyzers and Hall effect set up) used in this work. The authors also thank Prof. J.S. Speck for a careful review of the manuscript.
.